\documentclass{article}
\usepackage{graphicx} % Required for inserting images
\usepackage{amsmath} % for align environment
\usepackage{comment}
\usepackage{authblk}
\usepackage[numbers]{natbib}

\title{Road-Width-Aware Network Optimisation for Bike Lane Planning}

\author[1]{Riccardo Basilone}%\email{riccardoroberto.basilone@uniroma1.it}
\author[2,3]{Matteo Bruno}
\author[2,3,4]{Hygor Piaget Monteiro Melo}
\author[1]{Michele Avalle}
\author[1,2,3,5]{Vittorio Loreto}

\small{
\affil[1]{Sapienza Univ. of Rome, Physics Dept, Piazzale A. Moro, 2, 00185, Rome, Italy}}
\affil[2]{Sony Computer Science Laboratories - Rome, Joint Initiative CREF-SONY, Centro Ricerche Enrico Fermi, Via Panisperna 89/A, 00184, Rome, Italy}
\affil[3]{Centro Ricerche Enrico Fermi (CREF), Via Panisperna 89/A, 00184, Rome, Italy}
\affil[4]{Postgraduate Program in Applied Informatics, University of Fortaleza, 60811-905, Fortaleza, Ceará, Brazil}
\affil[5]{Complexity Science Hub, Josefst\"{a}dter Strasse 39, A 1080 Vienna, Austria}
\begin{document}

\maketitle

\begin{abstract}

Active mobility is becoming an essential component of the green transition in modern cities. However, the challenge of designing an efficient network of protected bike lanes without disrupting existing road networks for motorised vehicles remains unsolved. This paper focuses on the specific case of Milan, using a network approach that considers street widths to optimise the placement of dedicated bike lanes at the edges of the network. 
Unlike other network approaches in this field, our method considers the actual shapes of the streets, which introduces a realistic aspect lacking in current studies. We used these data to simulate cycling networks that maximise connectivity while minimising the impact of bike lane placement on the drivable network. Our mixed simulation strategies optimise for edge betweenness and width.
Furthermore, we quantify the impact of dedicated bike lane infrastructure on the existing road network, demonstrating that it is feasible to create highly effective cycling networks with minimal disruption caused by lane width reductions. This paper illustrates how realistic cycling lanes can be simulated using road width data and discusses the challenges and benefits of moving beyond one-dimensional road data in network studies.

\end{abstract}

\section{Introduction}\label{sect: Introduction}

Reducing the amount of time people spend driving is widely recognised for its positive impact on pollution levels and overall quality of life~\cite{DeNazelle:2011ActiveTransport, Kent:2014Driving4Time, Nieuwenhuijsen:2016CarFreeCities, Prieto:2024ABC-of-Mobility, Rabl:2012BenefitsOfActiveTransport}. This reduction can be achieved both by decreasing the number of people driving and by alleviating traffic congestion on the streets. Both effects can be simultaneously obtained by encouraging alternatives to automobile transport, such as cycling~\cite{DeNazelle:2011ActiveTransport, Rabl:2012BenefitsOfActiveTransport}.\par  
Cities are key players in this regard, as daily urban trips are often suited for bicycle or e-bike travel.\par 
Efforts to move away from car dependency in cities are also part of a general effort to move towards a more human-centric urban framework~\cite{Rhoads:2021, Rhoads:2023}. Car-centric cities are on the opposite philosophy of urban life from proximity models, such as the ``15-minute city"~\cite{Bruno:2024universal, Marzolla:2024compact, 
%Moreno:2016-15minCity, 
Moreno:2021-15minCity2, hill2024cities}. \par
The concept of the 15-minute city is to rethink cities in such a way that citizens have access to essential services within a 15-minute trip with non-polluting means such as walking or cycling. Though not every city can hope to achieve the ideal 15-minute threshold due to structural characteristics such as a stark distinction between central and peripheral areas~\cite{Bruno:2024universal, Fanelli:2024core-periphery}, steps can be taken to improve urban accessibility, and the role of bicycles can help extend this concept to periurban and low-density areas. \par  
That is why many cities with poor cycling infrastructure are working on improving: urban science has shown that cycling uptake increases significantly when cyclists have safe, well-connected cycling routes at their disposal~\cite{Kraus:2020BikeLaneEffect}. This process is an effect of a more general social phenomenon known as ``induced demand"~\cite{Lee:1999inducedDemand}, that is, the demand for something rising because its supply has gone up, and not vice versa. \par
Note that just as induced demand predicts an increase in cyclists when infrastructure is built, it also tells us that some people may take up driving once they see improved traffic flow and decreased commuting times; still, studies have shown that introducing cycling lanes on roads can have little negative effect on the travel times of those roads~\cite{Nanayakkara:2022BikeLaneTraffic}.\par
Urban planning for the placement of new cycling lanes often acts at a localised, single street or neighbourhood level~\cite{CROW:2016design, Szell:2022}. However, recent studies have begun treating cycling lane placement as a complex network problem. For example, in Natera et al.~\cite{Natera:2020data-driven} and Szell et al.~\cite{Szell:2022}, the authors devise data-driven strategies to simulate networks with desirable characteristics for riders, such as connectivity, coverage and directness. They apply their framework to 62 cities worldwide. Single city case studies also exist, where a network is simulated with a parameter to balance between optimisation of safety or efficiency~\cite{Folco:2023Turin}. The models developed in these studies tend to yield good results on low projected budgets. \par 
These works show the importance of holistic, network-level approaches in two ways: firstly, by demonstrating how cities with poor infrastructure could allocate a small amount of resources to improve greatly; secondly, by showing that many synthetic optimised networks have close correspondence to existing networks in cities that are considered to have a good cycling infrastructure. \par
The aforementioned studies treat road networks as linear geometries. Data on road length is readily available for many cities and is relevant in most routing problems. On the other hand, we find that information on road width would be very important to consider, as it can tell us which roads are physically able to contain dedicated lanes, and which aren't. Furthermore, it is reasonable to expect the width of any dedicated lane to be relevant in assessing its impact on the existing road network: dedicated lanes for public transport can be several times wider than unprotected cycling lanes, for example~\cite{DM2001}. \par 
The main obstacle to adding width information to urban street network studies is that road width data is quite difficult to find. Open data sources such as OpenStreetMap~\cite{OpenStreetMap} are well compiled in many aspects, but road width is not one of them. The field of computer vision has early developments in automated extraction and segmentation of road geometry from satellite images~\cite{alshehhi2017hierarchical}, but publicly available data sets with road width data compiled are few and far between. \par 
This paper introduces a new realistic framework for using road width to simulate dedicated lane infrastructure. Using a data set from the city of Milan, we build a road network of the city, starting from the polygons of its drivable streets. This allows us to extract the width and length of each road segment. Next, we simulate the addition of dedicated lanes on top of existing roads, creating new subnets. We assign a score to roads based on their width and centrality (using the betweenness centrality network measure), which are proxies for competing characteristics: the most optimised cycling networks are built on high betweenness edges, but the ones with the least impact on cars are built on the widest edges. We simulate the introduction of protected lanes by selecting street segments based on a mixing parameter between these two quantities, and we assess the impact of the resulting synthetic networks on the existing drivable network. \par
Our results show how it is possible to effectively simulate dedicated networks on a whole city while incorporating road width and that the extent to which width is considered can significantly influence the resulting networks.
This study confirms and extends previous work, by adding an extra degree of realism to the simulated networks. We process polygonal data to obtain road graphs comprehensive of physical road characteristics, to then show that desirable network characteristics for cyclists can be in direct competition with ones that minimise automobile congestion. We also show that the impact of cycling networks on existing roads is minimal even while implementing two-way dedicated lanes 2.5 m wide. These findings serve as a great encouragement to build more cycling lanes with a holistic, network-based view in mind.

\section{Methods}\label{sect: methods}

In this section, we describe the gathering and preprocessing of data to create street networks that include width information. Then we detail the simulations we performed to implement bike lanes.

\subsection{Creating the road network}\label{subsect: road net creation}

\begin{figure}
    \centering
    \includegraphics[width=1\linewidth]{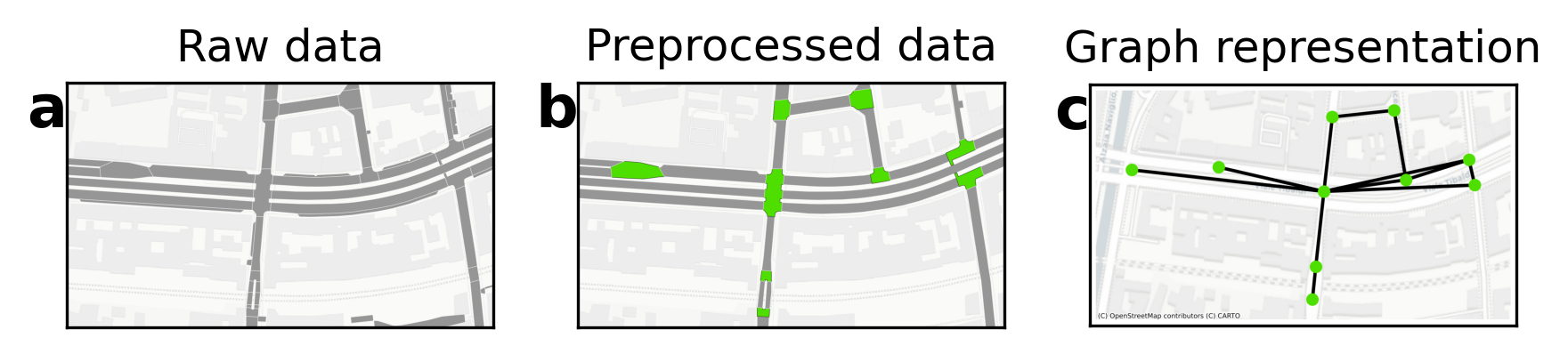}
    \caption{Data processing, from road polygon to graph. The figure shows a sample of the dataset. Starting from the road polygon of the city \textbf{(a)}, we classify sections of the road and remove all elements besides intersections and the roads connecting them \textbf{(b)}. Each intersection becomes a node of the network, while roads become edges, with attributes such as length and width \textbf{(c)}.
    }
    \label{fig:preprocessing}
\end{figure}

Our data set is part of a standard called \textit{Database Geotopografico}, made on a regional or municipal basis in Italy~\cite{DPCM2011, CatalogoDati:2015}. It contains contiguous polygons of the drivable area of Milan in 2020. Polygons are classified with relevant infrastructural information along with their geometries. 
\par
Figure~\ref{fig:preprocessing} summarises the processing stages from raw shapefile \textbf{(a)}, to the classified road polygons \textbf{(b)}, to the road graph \textbf{(c)}.

\subsubsection{Data processing}\label{subsubsect: data processing}

The full dataset also includes geometries that are not related to drivable roads, such as pedestrian areas and parking. Therefore we remove all polygons except for those classified as:
\begin{itemize}
    \item Main carriageway. Type ``01" in the data. 
    \item Structured traffic areas, \textit{i.e.} intersections, roundabouts and squares. Type ``02" in the data.
\end{itemize}
Type ``02" elements will become the nodes of the network, while type "01" ones will be the edges. \par 
Often, parking spaces on the side of the carriageway are part of the polygons, except when road curb and shape are significantly altered by their presence. When that is the case, they are distinguished into separate polygons, which we remove. We choose to exclude these polygons from our analysis because imagining their transformation into dedicated lanes would necessarily entail modifying the parts of the curb that encase them, which is more complicated than just reallocating existing road space from cars to bicycles (see Supplementary Material for an analysis of the effect of keeping these parking spaces).\par
After merging adjacent carriageway segments into one, the data is in the shape of Figure~\ref{fig:preprocessing} (panel b), with only structured traffic areas connected by main carriageways.

\subsubsection{Width and length calculation}\label{subsubsect: width calc}
    In order to calculate the length and width of a polygon in our data set, we approximate its shape to be rectangular. After computing area and perimeter numerically, we can write them as functions of length and width:
\begin{align*}
   \left\{
   \begin{array}{ll}
      A &= lw \\
      P &= 2l + 2w,
   \end{array}
   \right.
\end{align*}
where $A$ and $P$ are area and perimeter, while $l$ and $w$ are length and width.
The solutions to these equations are the estimated length and width of the polygon. Width can always be taken as the smaller of the two solutions, for our data set, as almost all road segments are longer than they are wide.
   
\subsubsection{Graph creation from polygons}\label{subsubsect: graph from polygon}
Once the road polygons have been merged where needed, and have length and width assigned, we create the road graph. To do this, we assign a node to each intersection, and an edge to each road. Edges are attached to nodes when they share a boundary. Nodes sharing boundaries with the same edge are connected.
This gives us an undirected network of Milan's drivable roads, with information on length and width. \par 
We also calculate the betweenness centrality of each edge. The betweenness of each edge $e$ can be written as
\begin{align*}
b(e) = \sum_{i,j \in V} \frac{\sigma(i,j|e)}{\sigma(i,j)},
\end{align*}
where $i,j$ are nodes in the set of nodes $V$, $\sigma(i,j|e)$ is the number of shortest paths between them that pass through $e$, and $\sigma(i,j)$ is the number of all shortest paths between $i$ and $j$. We weigh all shortest paths by the length of road segments.

\subsection{Simulations}\label{subsect: Simulations}
We simulate the creation of dedicated cycling lanes on roads by building subgraphs of the drivable graph. To select which edges will be part of a simulation, we assign scores to each edge, based on its width and betweenness. 
\subsubsection{Assigning scores to edges}
The score assigned to each edge is based on their width ($w$) and betweenness ($b$). To make the two quantities comparable, we compute their percentile ranks
\begin{align*}
R_w(e) = \left(1-\frac{Rank_w(e)}{N_e}\right) \times 100\\
R_b(e) = \left(1-\frac{Rank_b(e)}{N_e}\right) \times 100,
\end{align*}
where $Rank_{w/b}(e)$ is the ranking of edge $e$ in width/betweenness, and $N_e$ is the total number of edges. This is the probability of finding an edge ranked below $Rank(e)$ if the data's distribution corresponds to the observed one. This mapping assigns a value between 0 and 1 to each edge, that is only a function of its rank compared to other edges, and not the magnitude of the width or betweenness. \par
Finally, we can assign a score to each edge $e$ as
\begin{align}\label{eq: score}
    s(e) = \alpha R_w(e) + (1-\alpha)R_b(e), 
\end{align}
where $\alpha \in [0,1]$  is a mixing parameter that varies over simulations. After choosing a value for $\alpha$, we rank edges by score. \par
Lower values of $\alpha$ increase the importance of edge betweenness, and higher ones increase the importance of edge width. As we discuss in subsequent sections of the paper, subgraphs created from edges with high betweenness tend to have better connectivity, whereas ones created from wider edges tend to have a smaller impact on the flow of automobiles in the original network. 

\subsubsection{Simulation procedure}
As a first step we set the width of the dedicated lanes, $w$. We exclude all edges whose widths are smaller than some threshold $\tau$. The threshold is the width of the dedicated lane plus the width we keep available for car drivers. For example, to simulate bike lanes with a width $w$ of 2.5 m, the threshold is set to 6 m, so as to have at least 3.5 m available for cars after the simulation, in line with guidelines from the Italian \textit{Codice della Strada} (Highway Code)~\cite{DM2001}. 

For the second step we choose an $\alpha$ between 0 and 1, and rank edges by their score calculated in Equation~\ref{eq: score}. This determines whether betweenness or width will contribute more to the score of each edge. \par 
The next step is to define a length budget in kilometres, $L$: during a simulation, edges are added to the subgraph until the length budget is reached. \par 
At the end of the simulation, we update the widths of the edges of the road graph: edges where cycling lanes were placed will have $w$ subtracted from their width, while the others remain untouched. \par %The edges should be removed instead... vabbè
We performed simulations with parameters in the following ranges:
\begin{itemize}
    \item $w$ from 1 m to 5 m, with a 0.5 m step
    \item $\tau$ is always $w+3.5$, as prescribed by Italian law on minimum car lane width.
    \item $\alpha \in \{0, 0.1, 0.3, 0.5, 0.7, 0.9, 1\}$ 
    \item $L$ from 1 km to 20 km with a 1 km step; from 20 km to 200 km with a 10 km step; from 200 km to 1000 km with a 40 km step.
\end{itemize}

\section{Results}
In this section we show the results of our simulations and analyse the characteristics of the drivable network along with the simulated networks. All results discussed are relative to two-way cycling lanes of 2.5 metres of width.

\subsection{Drivable network characteristics} 
\begin{figure}
    \centering
    \includegraphics[width=1\linewidth]{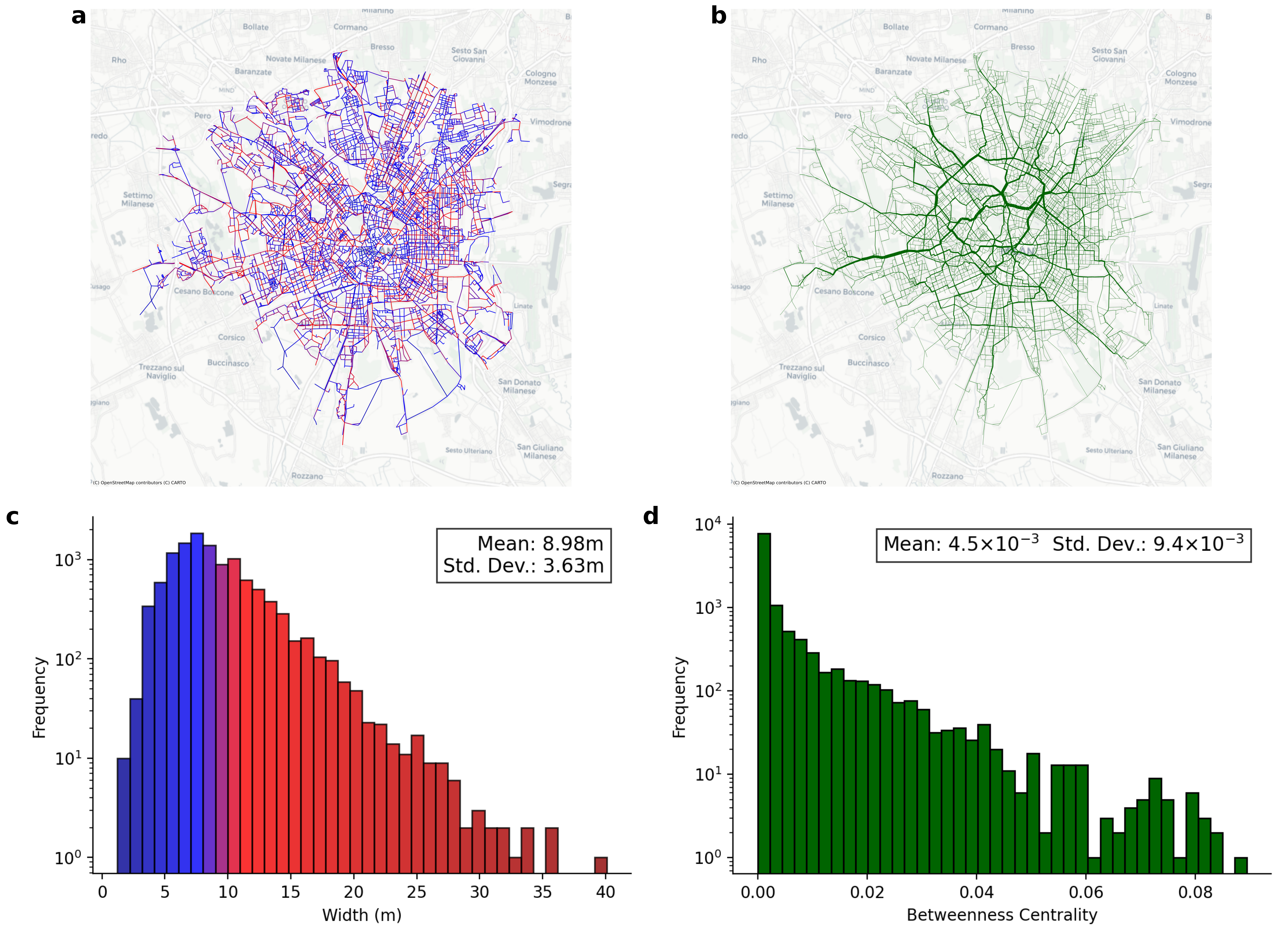}
    \caption{Characteristics of the driving network. \textbf{(a,c):} Color map and histogram of the width of the edges. Edges narrower than the mean width (8.98m) are shown in blue, while wider edges are in red. \textbf{(b)} Betweenness map of the edges. Edges with higher betweenness are represented as larger. This plot is useful to see Milan's arterial roads. \textbf{(d)} Histogram of the edge betweenness. Most edges of the network have low betweenness, meaning that a limited number of streets appear on many shortest paths in the city.}
    \label{fig:drivenet}
\end{figure}

The shape and characteristics of the drivable network are shown in Figure~\ref{fig:drivenet}. The 11323 edges in the network have widths spanning from 3 to 40 metres, with an average width of approximately 9 metres, and a total length of 1743 km. The distribution peaked around the average. Considering the minimum mandated lane width of 2.75 to 3.5 m for Italian streets, this means that the average Milanese street has at least two lanes. \par
On the other hand, the edge betweenness distribution shows a peak near zero, with a few tens of streets acting as Milan's arterial routes. These streets are the ones selected the earliest in simulations at low values of $\alpha$.
the top 10\% of streets by betweenness are 162 km long in total. \par
Although it is natural to suspect that wider streets of the city may be more important and, therefore, have higher betweenness, no significant correlation exists between street width and betweenness centrality (see Supplementary Material). %Could be put here below the two histograms

 \subsection{Simulation results}
 The results of our simulations do not take into account any existing cycling infrastructure already present in Milan. This is because we aim to demonstrate the shape and characteristics of networks built from scratch, taking space from drivable roads.

 \subsubsection{240 km budget: a realistic example} 
 
 \begin{figure}
    \centering
    \includegraphics[width=1\linewidth]{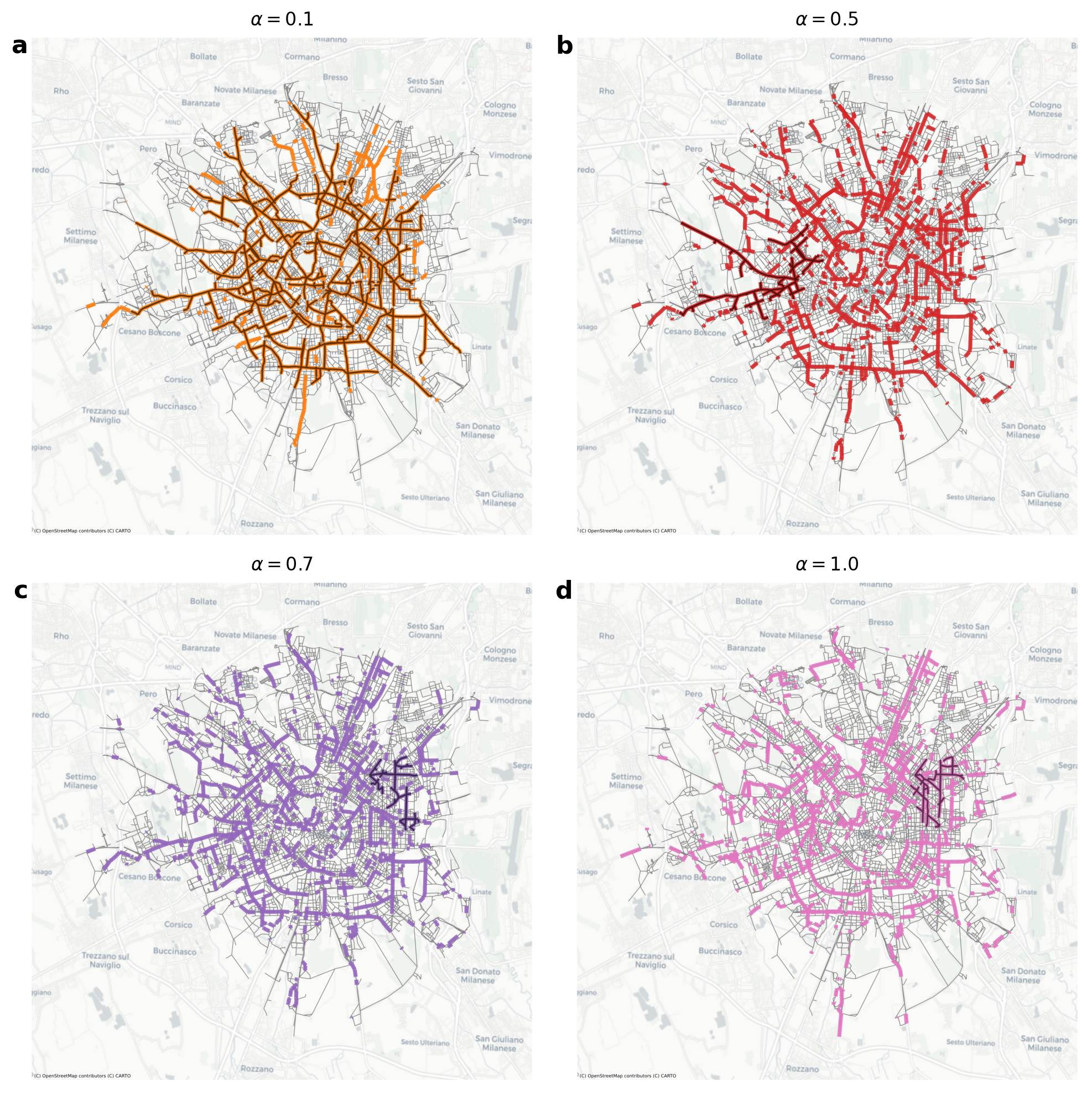}
    \caption{\textbf{(a)-(d)} visualisation of the simulation results for four different values of $\alpha$, with a budget of $L=240km $: the simulated cycling network is plotted in a light colour, while the largest connected component (LCC) of each simulation is highlighted in a more intense hue. Lower values of $\alpha$ are simulations that privilege betweenness. The size of the LCC at $\alpha = 0.1$ suggests that the network percolates.}
    \label{fig:results maps}
\end{figure}

Figure~\ref{fig:results maps} shows four simulated networks at a fixed budget of $L = 240$ km and four different values of $\alpha$. They are similar in length to Milan's actual cycling infrastructure network, which is a little over 200 km~\cite{AMAT:2022ciclabili}. \par 
A giant connected component is visible in Figure~\ref{fig:results maps} \textbf{(a)}, where $\alpha =0.1$, showing that a percolation transition has occurred. The same cannot be said at this budget for larger values of $\alpha$. This was expected, as values closer to 1 are simulations that privilege road width over betweenness and, therefore, over network connectivity. See Supplementary Material for more images.
 \subsubsection{Cycling network connectivity measures
 }
\begin{figure}
    \centering
    \includegraphics[width=1\linewidth]{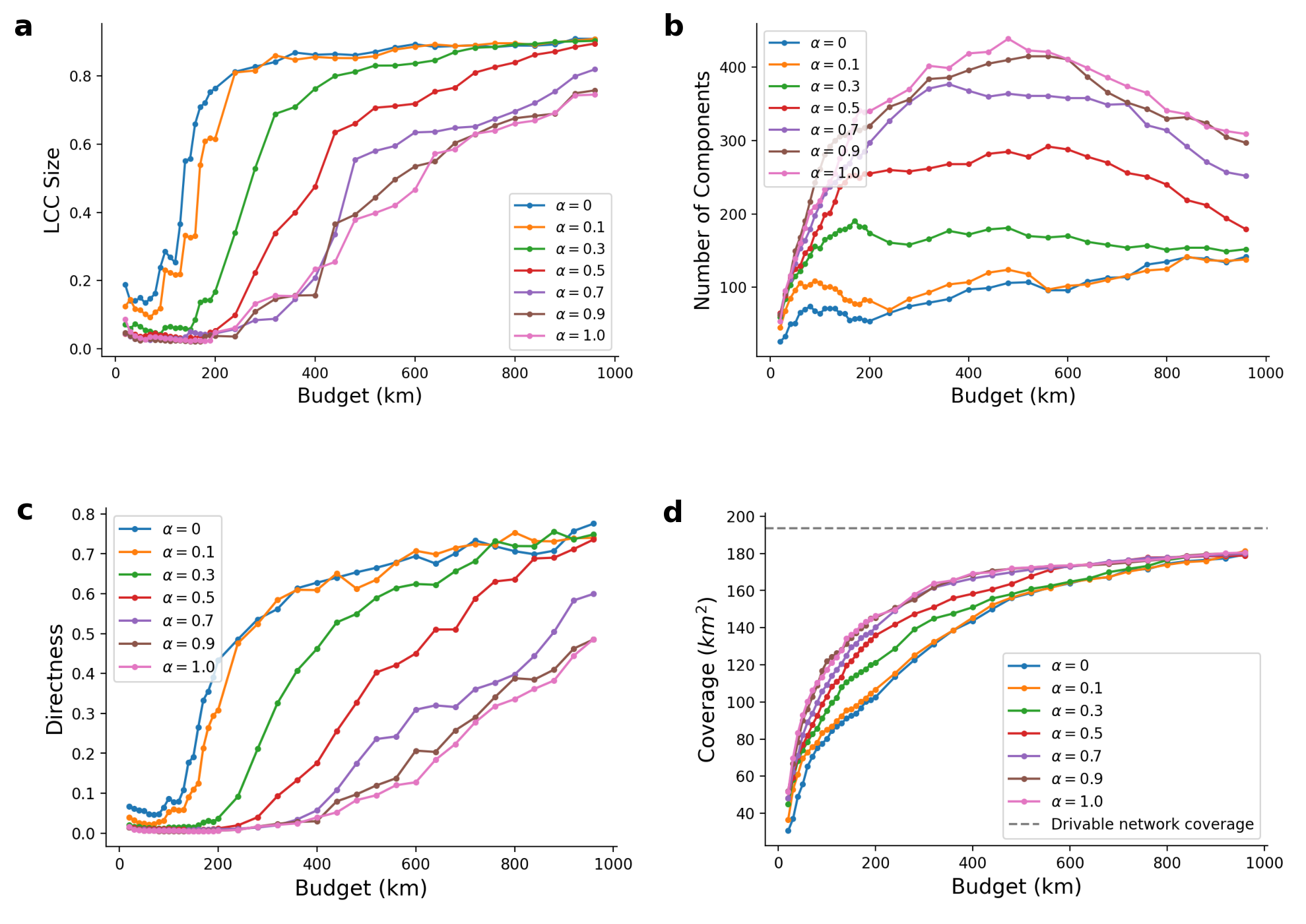}
    \caption{\textbf{(a),(b)} Size of the largest connected component and number of components of the simulated networks, for all $\alpha$. Lower values of $\alpha$ result in early growth of the largest connected component, as in a percolation process. The total number of components also stays lower. \textbf{(c), (d)} Evaluation metrics of the simulated networks, for all $\alpha$: Directness measures the ratio between the length of the shortest driving path vs the length of the shortest biking path between all possible origins and destinations. Higher values mean fewer detours during a trip. Coverage is measured as in~\cite{Natera:2020data-driven}: it is the area of the simulated network, endowed with a 500 m buffer in every direction. The coverage of the drivable network is represented as a grey, dashed line. Higher values of $\alpha$ result in more fragmented networks, that cover a larger area than dense, well-connected ones.}
    \label{fig:results graphs}
\end{figure}

 We study percolation by monitoring the size of the largest connected component (LCC) of the simulated network, along with the total number of components present.
 Here, by percolation, we refer to the process by which a large connected component—whose size is comparable to that of the total network—emerges very rapidly as the network grows. In simple terms, a subgraph that percolates is one with a large region of nodes that have paths connecting them.
 In Figure~\ref{fig:results graphs} \textbf{(a)}, the size of the LCC increases with budget for all values of $\alpha$, but critical percolation behaviour seems to only appear for $\alpha \le 0.3$. As $\alpha$ grows, edge width gains more importance compared to betweenness, so the LCC grows in a more linear fashion. \par 
 In Figure~\ref{fig:results graphs} \textbf{(b)}, the number of total network components has small fluctuations for lower values of $\alpha$, showing that new edges added to the network tend to be attached to already existing ones. For higher values of $\alpha$, the number of components peaks at a budget of around 500 km, indicating that it seems to become statistically more probable that new edges are attached to existing network components at higher budgets than that. \par 

 \subsubsection{Directness and coverage}
 We further evaluate our results by measuring the directness and coverage of our simulated networks, for all $\alpha$, in a similar way to Natera et al.~\cite{Natera:2020data-driven}. We define directness as the average ratio of shortest distances between two nodes of the network by car or by bike, as 
 \begin{align*}
     \Delta = \langle \frac{\delta^{car}_{ij}}{\delta^{bike}_{ij}}\rangle_{ij},
 \end{align*}
 Where $\delta_{ij}$ is the shortest distance between a pair of nodes when travelling on the bike or car network, and $\langle .\rangle_{ij}$ is the average over all the origin-destination node pairs. Since the cycling network is always a subgraph of the drivable network, $\delta^{bike}_{ij}\ge \delta^{car}_{ij}$. When two nodes are not connected by a bike path, the ratio $\frac{\delta^{car}_{ij}}{\delta^{bike}_{ij}}$ is set to 0.\par
 Figure~\ref{fig:results graphs} \textbf{(c)} shows directness of the simulated networks as a function of budget (in km), for all $\alpha$. 
 As expected, lower values of $\alpha$ yield networks with higher directness, thanks to increased connectivity. \par 
Coverage is measured as defined in Natera et al.~\cite{Natera:2020data-driven}: it is the area of the simulated network, endowed with a 500 m buffer in every direction, which represents a proxy for a 15-minute walking distance from the network. \par 
Higher values of $\alpha$ result in more fragmented networks that cover a larger area than dense, well-connected ones (Figure~\ref{fig:results graphs} \textbf{(d)}). As the networks grow in kilometres, behaviour at different values of $\alpha$ becomes more and more similar, approaching a maximum of 193 km$^2$, which is the coverage of the drivable network. \par 
High scores in these two evaluation metrics correspond to desirable, complementary characteristics of the network: directness indicates the number of detours away from the network needed to reach a destination, and coverage indicates the number of destinations that are reachable.

\subsubsection{Impact on Drivable Network}\label{subsubsect: impact on cars}
 Studying the direct effects of cycling lanes on existing roads would require simulations that go beyond the scope of this structure-centred work. However, studies have shown that decreasing the number of two-way edges in a network significantly increases congestion and impacts the existence of paths between pairs of nodes~\cite{Carmona:2020OPC,melo2022impact}.\par 
 With this in mind, we quantify the impact on the car network of our simulations by calculating the percentage of streets going from two-way to one-way. As per the Italian \textit{Codice della strada}~\cite{DM2001}, lanes in primary roads must be at least 3.5 m wide, while lanes in residential roads must be at least 2.75 m wide. We assume that all edges wider than 7 m are two-way, so all edges that become narrower than 7 m after a simulation have become one-way. Note that this assumption is conservative, in that it maximises the amount of edges counted as impacted by our simulations. \par
\begin{figure}
     \centering
     \includegraphics[width=1\linewidth]{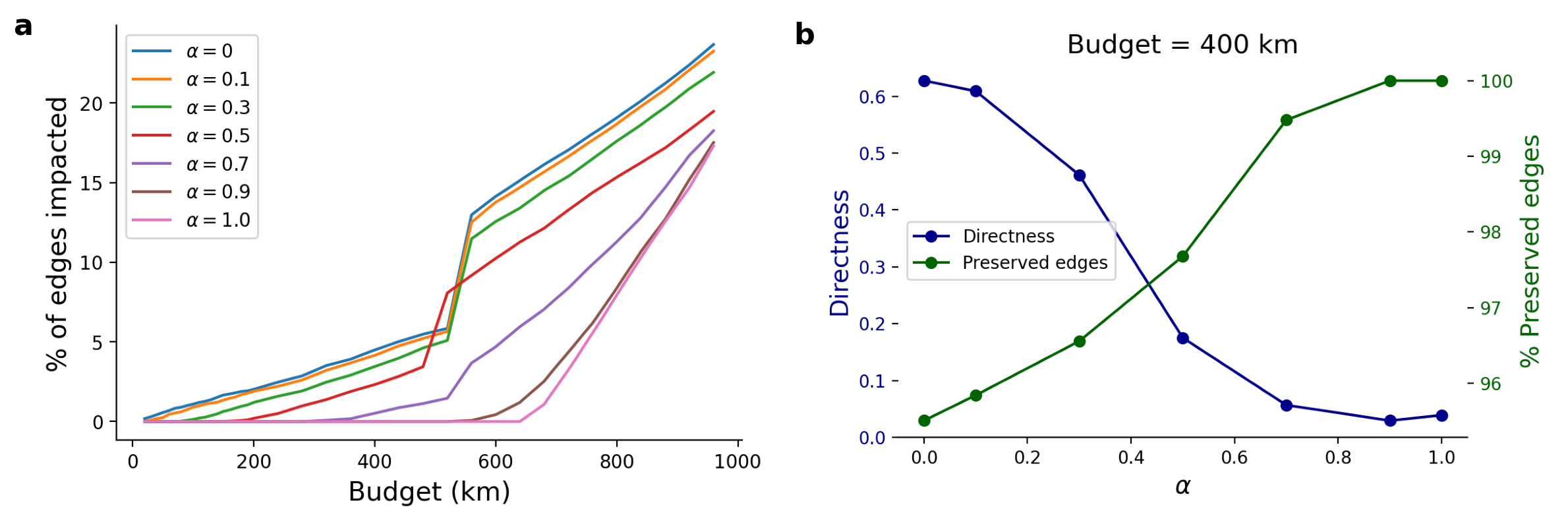}
     \caption{\textbf{(a)} The percentage of lanes going from two way to one way after simulations, for all values of $\alpha$. 
     \textbf{(b)} A comparison between directness and \% of impacted lanes, at a fixed budget. Lower values of $\alpha$ make for more direct bike networks but preserve a smaller percentage of two-way edges.
     }
     \label{fig:car impact}
 \end{figure}

  Figure~\ref{fig:car impact} \textbf{(a)} shows the percentage of impacted lanes on the drivable network as a function of budget. We see that growing values of $\alpha$ allow simulations to preserve more edges, even at higher budgets. This is in line with the fact that higher values of $\alpha$ privilege edge width compared to betweenness. \par 
  Figure ~\ref{fig:car impact} \textbf{(b)} compares directness to the percentage of preserved edges, as $\alpha$ varies. The plot highlights the competition between a highly direct cycling network and one with a small impact on drivers. With 400 km of cycling networks, for $\alpha = 0.3$, the directness of the network is near 0.5, while over 96\% of two-way edges are preserved. See Supplementary Material for comparisons of other budgets.
 
\section{Discussion}

In this study, we advance the current state of the art by extracting and utilising road width information to simulate the creation of realistic dedicated cycling lane networks in Milan. We assess their effectiveness and impact on the existing road network. Our work is part of a broader effort to better understand the relationship between urban automobile transportation and its alternatives, emphasising the importance of properly distributing road space.\par
Firstly, we developed a strategy to consider the width and betweenness centralities when evaluating whether or not a given road is suitable to host a cycling lane. We also showed that these two edge attributes contribute to complementary characteristics of the emerging networks --- connectivity and impact on the drivable network. \par 
We mixed the width and betweenness attributes through a parameter $\alpha$ and showed different realisations of networks at length similar to that of Milan's actual cycling network, that is 240 km (Figure~\ref{fig:results maps}).\par
Networks created at low $\alpha$, favouring betweenness, exhibited percolating behaviour with a small number of components and a very large connected component. These networks also had higher directness than those at higher $\alpha$, where road widths are prioritised. Conversely, low $\alpha$ networks had less coverage than high $\alpha$ ones built on wider roads because they are denser and less fragmented. 
Finally, we evaluated the impact of our simulations on the driving network by calculating the percentage of two-way roads that became one-way after simulations (Figure~\ref{fig:car impact} \textbf{(a)}). High $\alpha$ strategies can reach budgets of over 500 km before having any relevant impact on the drivable network, but they will suffer significantly in directness compared to lower $\alpha$ networks. On the other hand, low $\alpha$ networks can achieve a directness greater than 0.5 while impacting less than 5\% of lanes with less than a 500 km budget. This evidence would be an excellent result for a city like Milan, which currently boasts about 200 km of cycling infrastructure for its almost 2000 km road. \par
Drawing some insight from our results, the best choice of $\alpha$ depends on the available budget. If there is little available budget compared to the total length of its roads, it makes sense to invest in a low $\alpha$ network. This will guarantee a high degree of connectivity by creating a sizeable connected component on the city's arterial streets. However, the impact on the driving network will not be negligible. \par
Conversely, if a city foresees a long-term project with a high budget, it may prefer higher values of $\alpha$. The network will be fragmented initially, but a giant connected component will eventually emerge, allowing the network to reach the connectivity and directness of its lower $\alpha$ counterparts while having a significantly smaller impact on the drivable network. \par 
This work does not aim to prescribe a specific best value for $\alpha$, so much as to show how adding a fundamental degree of realism, such as road width, to existing network models can significantly change how we interpret their predictions. \par
The evaluation metrics used for the bike network—directness and coverage—provide valuable insights, but they also have limitations. The directness metric is calculated based on the assumption that cyclists always use dedicated lanes. This represents a worst-case scenario and likely underestimates the actual usability, as cyclists may occasionally leave the network. On the other hand, the coverage metric indicates the network's reach, but it does not consider factors like population density or the distribution of essential urban amenities. Additionally, the assessment of the impact on the car network is based solely on the percentage of lost potential two-way streets, whereas the real impact is likely to be much more complex.\par

In this study, we concentrated on cycling infrastructure at a city-wide level. However, the methods we developed can also be applied to protected lanes of any width, such as bus lanes, at more localised scales if needed.\par
Future work could focus on refining evaluation metrics to include factors such as the population served, accessibility to key destinations, and the real-world routing behaviours of cyclists. Additionally, the impact of the new bike lanes on traffic could be estimated more accurately by simulating traffic in the city before and after the interventions. Furthermore, the existing bike lane networks could be compared to proposed routes to estimate the investment needed to enhance the directness of the network. Lastly, expanding this research to include more cities—such as by using satellite data to identify the actual shapes of streets—could broaden the scope of the current study on a more global scale.

\section{Conclusion}

In this study, we introduced a new framework for integrating road width data into the planning of dedicated cycling networks, using the city of Milan as a case study. By incorporating road width and betweenness centrality into the optimisation process, we examined the feasibility of developing extensive and well-connected cycling networks while minimally disrupting existing vehicular traffic. Our simulations indicated that even with the implementation of two-way cycling lanes that are 2.5 meters wide, less than 5\% of streets would need to be converted from two-lane roads to single-lane roads. This finding highlights the efficiency of such interventions.

Our findings revealed the trade-offs between the connectivity of implemented bike networks and the effects of new infrastructure on the drivable network. Networks optimised for betweenness centrality provide higher connectivity and direct routes but can significantly impact car traffic. Conversely, networks optimised for road width may be more fragmented but are less disruptive to existing traffic. This relationship gives urban planners flexible strategies based on their objectives. Bold policies could focus on creating a better and more direct bike network by prioritising betweenness centrality. On the other hand, more conservative policies might initiate a sustainable mobility transition by first establishing protected bike lanes on wider roads, thereby minimising traffic disruption and reducing skepticism.

This study highlights the significant potential of data-driven approaches to planning realistic urban cycling infrastructure. By using data on the shape and width of streets, we demonstrated that extensive and effective cycling networks can be developed with minimal disruption to existing roadways. Therefore, we advocate for the expansion of cycling infrastructure through policies that can lead to significant improvements while having a minimal impact on car traffic. This approach helps alleviate citizen concerns and facilitates a smoother, more widely accepted transition to sustainable mobility.

\section{Acknowledgements}
The authors declare no competing interests.

\newpage
	\bibliographystyle{unsrtnat}
	\bibliography{bibliography}

\begin{thebibliography}{27}
\providecommand{\natexlab}[1]{#1}
\providecommand{\url}[1]{\texttt{#1}}
\expandafter\ifx\csname urlstyle\endcsname\relax
  \providecommand{\doi}[1]{doi: #1}\else
  \providecommand{\doi}{doi: \begingroup \urlstyle{rm}\Url}\fi

\bibitem[De~Nazelle et~al.(2011)De~Nazelle, Nieuwenhuijsen, Ant{\'o}, Brauer, Briggs, Braun-Fahrlander, Cavill, Cooper, Desqueyroux, Fruin, et~al.]{DeNazelle:2011ActiveTransport}
Audrey De~Nazelle, Mark~J Nieuwenhuijsen, Josep~M Ant{\'o}, Michael Brauer, David Briggs, Charlotte Braun-Fahrlander, Nick Cavill, Ashley~R Cooper, H{\'e}l{\`e}ne Desqueyroux, Scott Fruin, et~al.
\newblock Improving health through policies that promote active travel: a review of evidence to support integrated health impact assessment.
\newblock \emph{Environment international}, 37\penalty0 (4):\penalty0 766--777, 2011.

\bibitem[Kent(2014)]{Kent:2014Driving4Time}
Jennifer~L Kent.
\newblock Driving to save time or saving time to drive? the enduring appeal of the private car.
\newblock \emph{Transportation research part A: policy and practice}, 65:\penalty0 103--115, 2014.

\bibitem[Nieuwenhuijsen and Khreis(2016)]{Nieuwenhuijsen:2016CarFreeCities}
Mark~J Nieuwenhuijsen and Haneen Khreis.
\newblock Car free cities: Pathway to healthy urban living.
\newblock \emph{Environment international}, 94:\penalty0 251--262, 2016.

\bibitem[Prieto-Curiel and Ospina(2024)]{Prieto:2024ABC-of-Mobility}
Rafael Prieto-Curiel and Juan~P Ospina.
\newblock The abc of mobility.
\newblock \emph{Environment International}, 185:\penalty0 108541, 2024.

\bibitem[Rabl and De~Nazelle(2012)]{Rabl:2012BenefitsOfActiveTransport}
Ari Rabl and Audrey De~Nazelle.
\newblock Benefits of shift from car to active transport.
\newblock \emph{Transport policy}, 19\penalty0 (1):\penalty0 121--131, 2012.

\bibitem[Rhoads et~al.(2021)Rhoads, Sol{\'e}-Ribalta, Gonz{\'a}lez, and Borge-Holthoefer]{Rhoads:2021}
Daniel Rhoads, Albert Sol{\'e}-Ribalta, Marta~C Gonz{\'a}lez, and Javier Borge-Holthoefer.
\newblock A sustainable strategy for open streets in (post) pandemic cities.
\newblock \emph{Communications Physics}, 4\penalty0 (1):\penalty0 183, 2021.

\bibitem[Rhoads et~al.(2023)Rhoads, Sol{\'e}-Ribalta, and Borge-Holthoefer]{Rhoads:2023}
Daniel Rhoads, Albert Sol{\'e}-Ribalta, and Javier Borge-Holthoefer.
\newblock The inclusive 15-minute city: Walkability analysis with sidewalk networks.
\newblock \emph{Computers, Environment and Urban Systems}, 100:\penalty0 101936, 2023.

\bibitem[Bruno et~al.(2024)Bruno, Monteiro~Melo, Campanelli, and Loreto]{Bruno:2024universal}
Matteo Bruno, Hygor~Piaget Monteiro~Melo, Bruno Campanelli, and Vittorio Loreto.
\newblock A universal framework for inclusive 15-minute cities.
\newblock \emph{Nature Cities}, 1\penalty0 (10):\penalty0 633--641, 2024.

\bibitem[Marzolla et~al.(2024)Marzolla, Bruno, Melo, and Loreto]{Marzolla:2024compact}
Francesco Marzolla, Matteo Bruno, Hygor Piaget~Monteiro Melo, and Vittorio Loreto.
\newblock Compact 15-minute cities are greener.
\newblock \emph{arXiv preprint arXiv:2409.01817}, 2024.

\bibitem[Moreno et~al.(2021)Moreno, Allam, Chabaud, Gall, and Pratlong]{Moreno:2021-15minCity2}
Carlos Moreno, Zaheer Allam, Didier Chabaud, Catherine Gall, and Florent Pratlong.
\newblock Introducing the “15-minute city”: Sustainability, resilience and place identity in future post-pandemic cities.
\newblock \emph{Smart cities}, 4\penalty0 (1):\penalty0 93--111, 2021.

\bibitem[Hill et~al.(2024)Hill, Bruno, Melo, Takeuchi, and Loreto]{hill2024cities}
Dan Hill, Matteo Bruno, Hygor~PM Melo, Yuichiro Takeuchi, and Vittorio Loreto.
\newblock Cities beyond proximity.
\newblock \emph{Philosophical Transactions A}, 382\penalty0 (2285):\penalty0 20240097, 2024.

\bibitem[Fanelli et~al.(2024)Fanelli, Melo, Bruno, and Loreto]{Fanelli:2024core-periphery}
Federica Fanelli, Hygor~PM Melo, Matteo Bruno, and Vittorio Loreto.
\newblock Revealing the core-periphery structure of cities.
\newblock \emph{arXiv preprint arXiv:2410.21133}, 2024.

\bibitem[Kraus and Koch(2020)]{Kraus:2020BikeLaneEffect}
Sebastian Kraus and Nicolas Koch.
\newblock Effect of pop-up bike lanes on cycling in european cities.
\newblock \emph{arXiv preprint arXiv:2008.05883}, 2020.

\bibitem[Lee~Jr et~al.(1999)Lee~Jr, Klein, and Camus]{Lee:1999inducedDemand}
Douglass~B Lee~Jr, Lisa~A Klein, and Gregorio Camus.
\newblock Induced traffic and induced demand.
\newblock \emph{Transportation Research Record}, 1659\penalty0 (1):\penalty0 68--75, 1999.

\bibitem[Nanayakkara et~al.(2022)Nanayakkara, Langenheim, Moser, and White]{Nanayakkara:2022BikeLaneTraffic}
Pivithuru~Kalpana Nanayakkara, Nano Langenheim, Irene Moser, and Marcus White.
\newblock Do safe bike lanes really slow down cars? a simulation-based approach to investigate the effect of retrofitting safe cycling lanes on vehicular traffic.
\newblock \emph{International journal of environmental research and public health}, 19\penalty0 (7):\penalty0 3818, 2022.

\bibitem[{CROW}(2016)]{CROW:2016design}
{CROW}.
\newblock \emph{Design Manual for Bicycle Traffic}.
\newblock CROW, Ede, Netherlands, 2016.
\newblock ISBN 978-90-6628-659-7.
\newblock URL \url{https://crowplatform.com/product/design-manual-for-bicycle-traffic/}.

\bibitem[Szell et~al.(2022)Szell, Mimar, Perlman, Ghoshal, and Sinatra]{Szell:2022}
Michael Szell, Sayat Mimar, Tyler Perlman, Gourab Ghoshal, and Roberta Sinatra.
\newblock Growing urban bicycle networks.
\newblock \emph{Scientific reports}, 12\penalty0 (1):\penalty0 6765, 2022.

\bibitem[Natera~Orozco et~al.(2020)Natera~Orozco, Battiston, I{\~n}iguez, and Szell]{Natera:2020data-driven}
Luis~Guillermo Natera~Orozco, Federico Battiston, Gerardo I{\~n}iguez, and Michael Szell.
\newblock Data-driven strategies for optimal bicycle network growth.
\newblock \emph{Royal society open science}, 7\penalty0 (12):\penalty0 201130, 2020.

\bibitem[Folco et~al.(2023)Folco, Gauvin, Tizzoni, and Szell]{Folco:2023Turin}
Pietro Folco, Laetitia Gauvin, Michele Tizzoni, and Michael Szell.
\newblock Data-driven micromobility network planning for demand and safety.
\newblock \emph{Environment and planning B: Urban analytics and city science}, 50\penalty0 (8):\penalty0 2087--2102, 2023.

\bibitem[{Ministero delle Infrastrutture e dei Trasporti}(2001)]{DM2001}
{Ministero delle Infrastrutture e dei Trasporti}.
\newblock {Decreto Ministeriale 5 Novembre 2001}, november 2001.
\newblock URL \url{https://www.gazzettaufficiale.it/eli/id/2002/01/04/01A13858/sg}.
\newblock Gazzetta Ufficiale della Repubblica Italiana, n. 273 del 23 novembre 2001.

\bibitem[{OpenStreetMap contributors}(2017)]{OpenStreetMap}
{OpenStreetMap contributors}.
\newblock {Planet dump retrieved from https://planet.osm.org }.
\newblock \url{ https://www.openstreetmap.org }, 2017.

\bibitem[Alshehhi and Marpu(2017)]{alshehhi2017hierarchical}
Rasha Alshehhi and Prashanth~Reddy Marpu.
\newblock Hierarchical graph-based segmentation for extracting road networks from high-resolution satellite images.
\newblock \emph{ISPRS journal of photogrammetry and remote sensing}, 126:\penalty0 245--260, 2017.

\bibitem[{Presidenza del Consiglio dei Ministri}(2012)]{DPCM2011}
{Presidenza del Consiglio dei Ministri}.
\newblock Decreto 10 novembre 2011: Adozione del sistema di riferimento geodetico nazionale, 2012.
\newblock URL \url{https://www.gazzettaufficiale.it/eli/id/2012/02/27/12A01799/sg}.
\newblock (GU Serie Generale n.48 del 27-02-2012 - Suppl. Ordinario n. 37).

\bibitem[per~l'Italia digitale(2015)]{CatalogoDati:2015}
Agenzia per~l'Italia digitale.
\newblock Specifiche di contenuto per i database geotopografici, december 2015.
\newblock URL \url{https://geodati.gov.it/geoportale/images/Specifica\_GdL2\_09-05-2016.pdf}.

\bibitem[e~Territorio(2022)]{AMAT:2022ciclabili}
Agenzia Mobilità~Ambiente e~Territorio.
\newblock Report della mobilità milano 2022, 2022.

\bibitem[Carmona et~al.(2020)Carmona, De~Noronha, Moreira, Ara{\'u}jo, and Andrade~Jr]{Carmona:2020OPC}
HA~Carmona, AWT De~Noronha, AA~Moreira, NAM Ara{\'u}jo, and JS~Andrade~Jr.
\newblock Cracking urban mobility.
\newblock \emph{Physical Review Research}, 2\penalty0 (4):\penalty0 043132, 2020.

\bibitem[Melo et~al.(2022)Melo, Mota, Andrade~Jr, and Ara{\'u}jo]{melo2022impact}
Hygor~PM Melo, Diogo~P Mota, Jos{\'e}~S Andrade~Jr, and Nuno~AM Ara{\'u}jo.
\newblock Impact of one-way streets on the asymmetry of the shortest commuting routes.
\newblock \emph{Physical Review Research}, 4\penalty0 (2):\penalty0 023053, 2022.

\end{thebibliography}

\end{document}

\maketitle

\newpage

\section{Correlation between width and betweenness}
\begin{figure}
    \centering
    \includegraphics[width=1\linewidth]{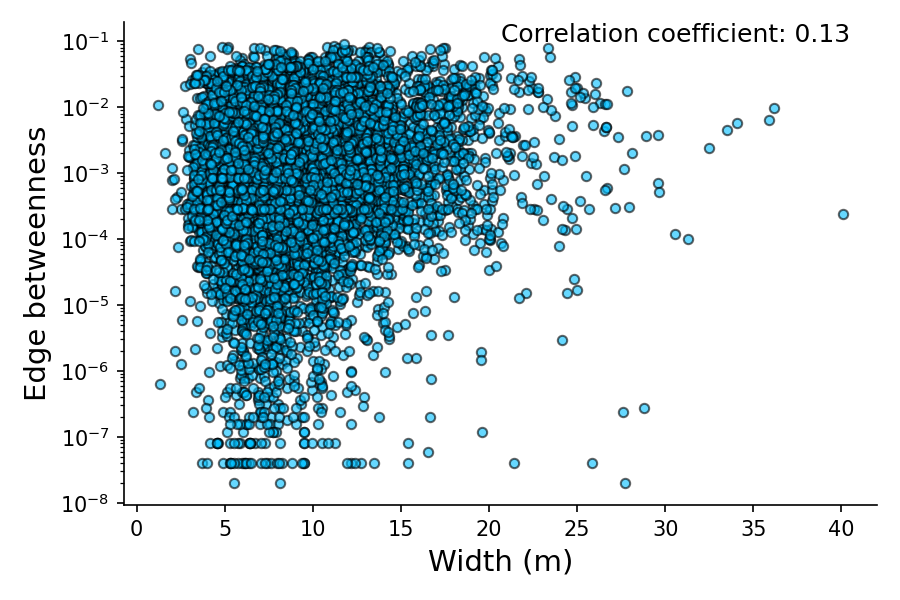}
    \caption{Width vs betweenness scatter plot. Correlation coefficient is $r = 0.13$, which is considered insignificant.}
    \label{fig:width vs bc scatter}
\end{figure}

We show the scatter plot of edge betweenness vs width in Figure \ref{fig:width vs bc scatter}. Pearson correlation coefficient r is not significant at 0.13.

\section{Data analysis tools}\label{subsubsect: OSMnx}
In recent years, the Python package OSMnx has proven to be extremely useful in studies regarding road networks and dedicated cycling lanes \cite{Boeing:2017OSMnx}. However, OSMnx creates its road networks from OpenStreetMap, which is severely lacking in regards to width data, only present on 0.4\% of roads worldwide \cite{OpenStreetMap}. %check this number again to be sure.
Integrating the polygonal data set we use into an OSMnx network is a far from trivial task, with lots of exceptions that must be handled manually, therefore in this work we build the road network with our own methodology. One prospect of future studies in this field is to create an efficient pipeline to integrate width data into OSMnx networks, so as to take advantage of its functionalities and possibly contribute to the open dataset.

\section{Network evolution for fixed $\alpha$}
\begin{figure}
    \centering
    \includegraphics[width=0.6\linewidth]{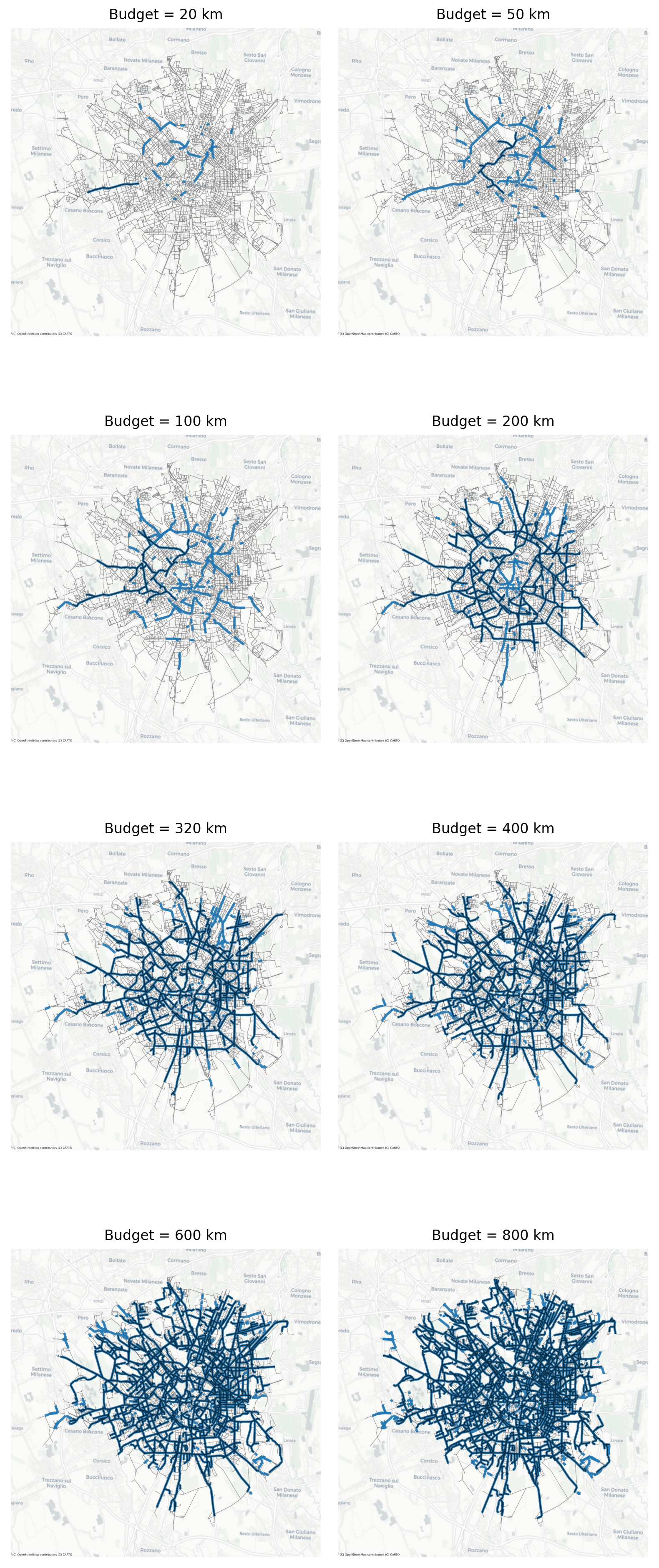}
    \caption{Network evolution at selected budgets for $\alpha = 0$. Percolation occurs almost immediately}
    \label{fig:net evolution alpha0}
\end{figure}

\begin{figure}
    \centering
    \includegraphics[width=0.6\linewidth]{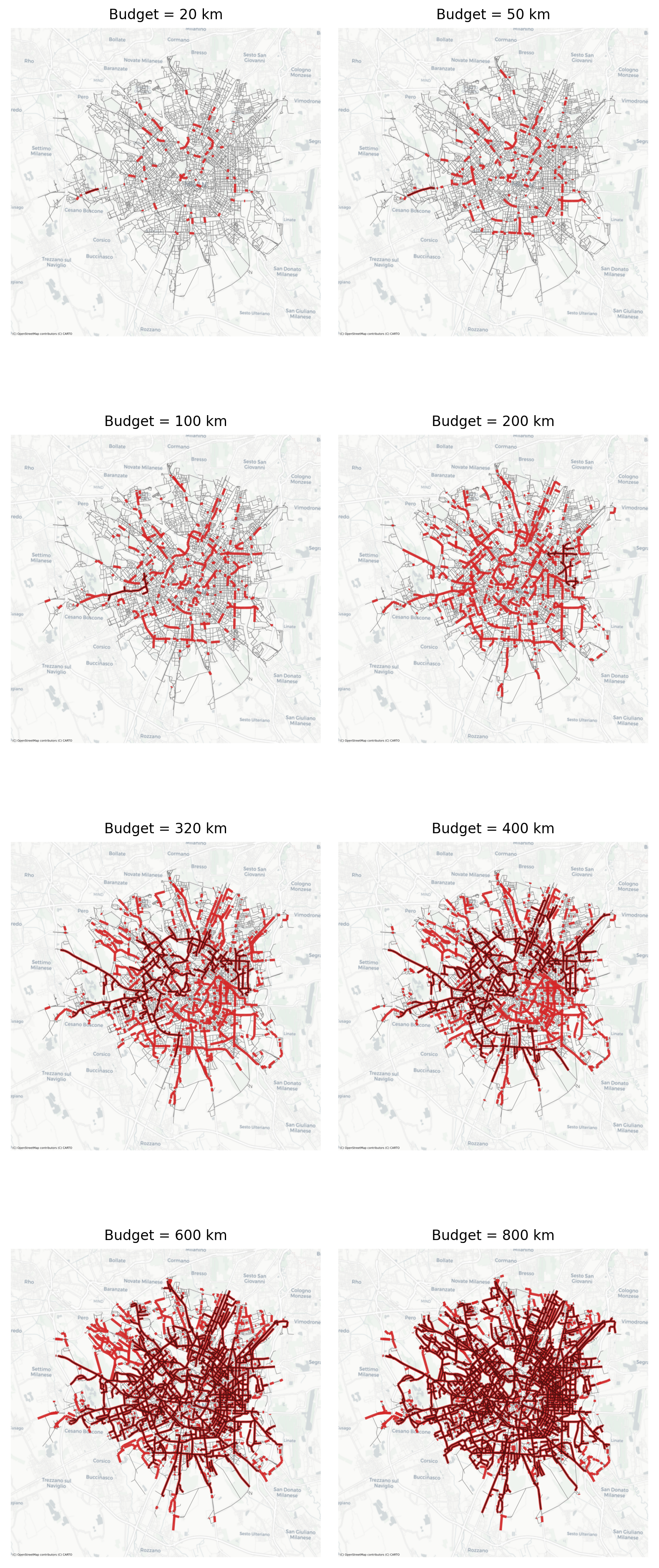}
    \caption{Network evolution at selected budgets for $\alpha = 0.5$. Percolation occurs very gradually, and late}
    \label{fig:net evolution alpha0.5}
\end{figure}

\begin{figure}
    \centering
    \includegraphics[width=0.6\linewidth]{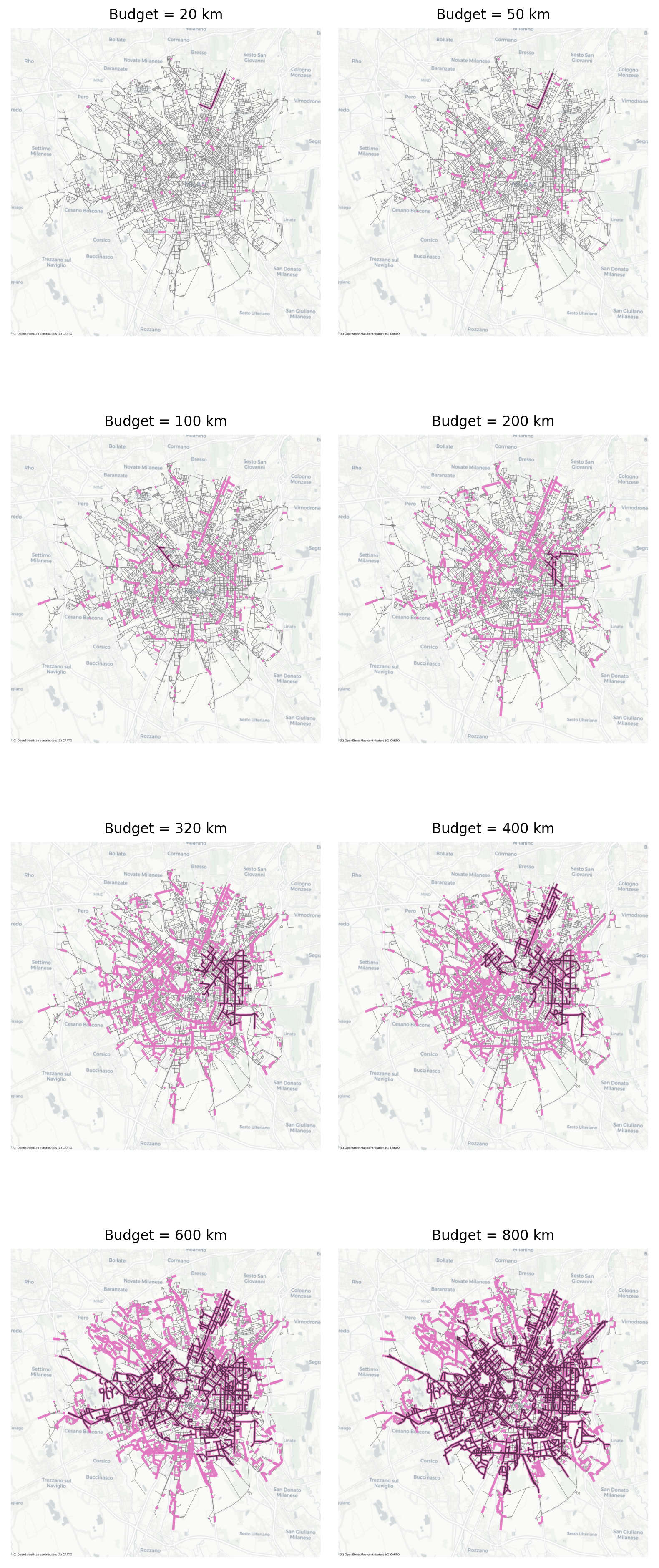}
    \caption{Network evolution at selected budgets for $\alpha = 1$. Percolation occurs very gradually, and late}
    \label{fig:net evolution alpha1}
\end{figure}

Figures \ref{fig:net evolution alpha0}, \ref{fig:net evolution alpha0.5}, \ref{fig:net evolution alpha1} show snapshots of the simulated network at selected budgets, for a given value of $\alpha$. Observing the progression of network growth can be useful to understand the effect of the value of $\alpha$. for $\alpha = 0$, the network grows from few components, and preserves a large connected component starting from low budgets. at $\alpha = 0.5$ width and betweenness are equally valued, so the network starts out more fragmented but eventually connects some of its larger components into one. at $\alpha = 1$ betweenness is ignored, and a giant component doesn't emerge until very high budgets are reached.

\section{Network snapshot at fixed budget for all $\alpha$} 
\begin{figure}
    \centering
    \includegraphics[width=0.6\linewidth]{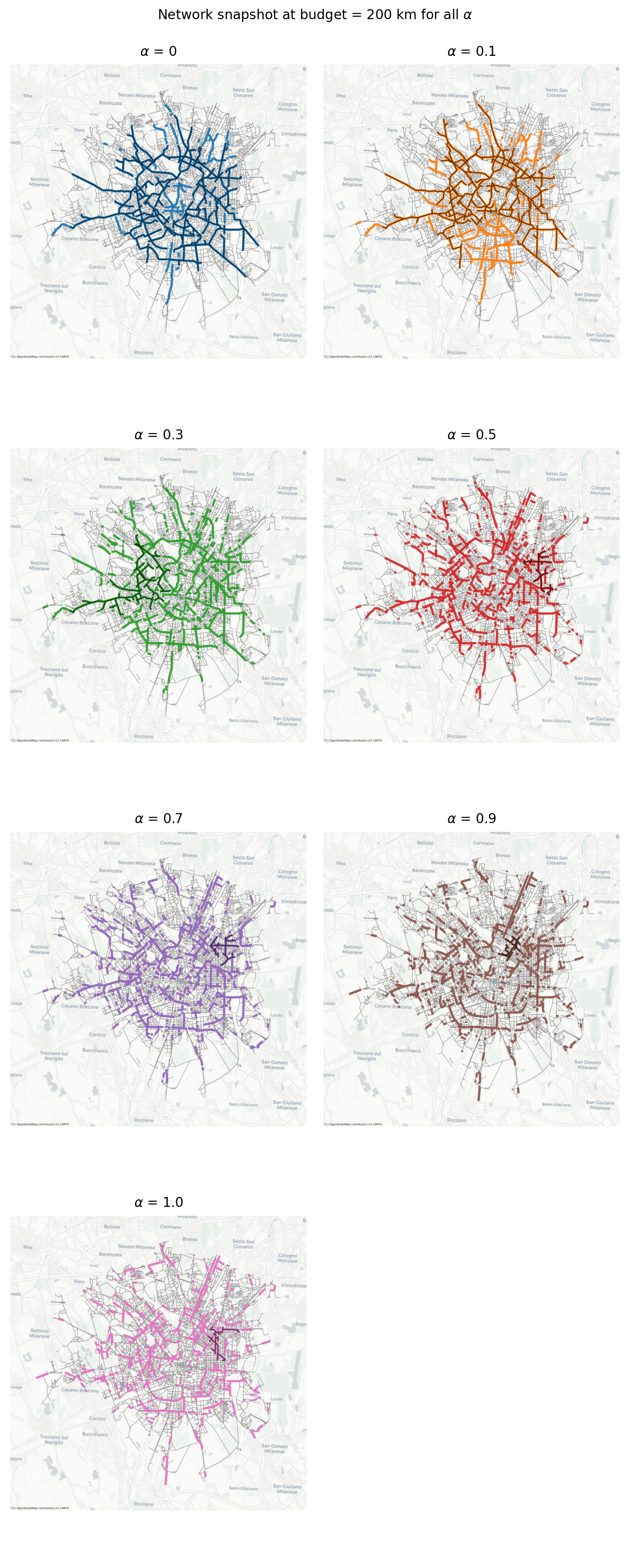}
    \caption{Simulation results for a budget of 200 km, for all $\alpha$.}
    \label{fig:net snapshot all alpha}
\end{figure}

Figure \ref{fig:net snapshot all alpha} shows a snapshot of all simulated networks at a budget of 200 km, for all $\alpha$. At this intermediate budget, only low $\alpha$ strategies have a visible largest connected component.

\section{Directness vs impacted car lanes over many budgets}
\begin{figure}
    \centering
    \includegraphics[width=0.9\linewidth]{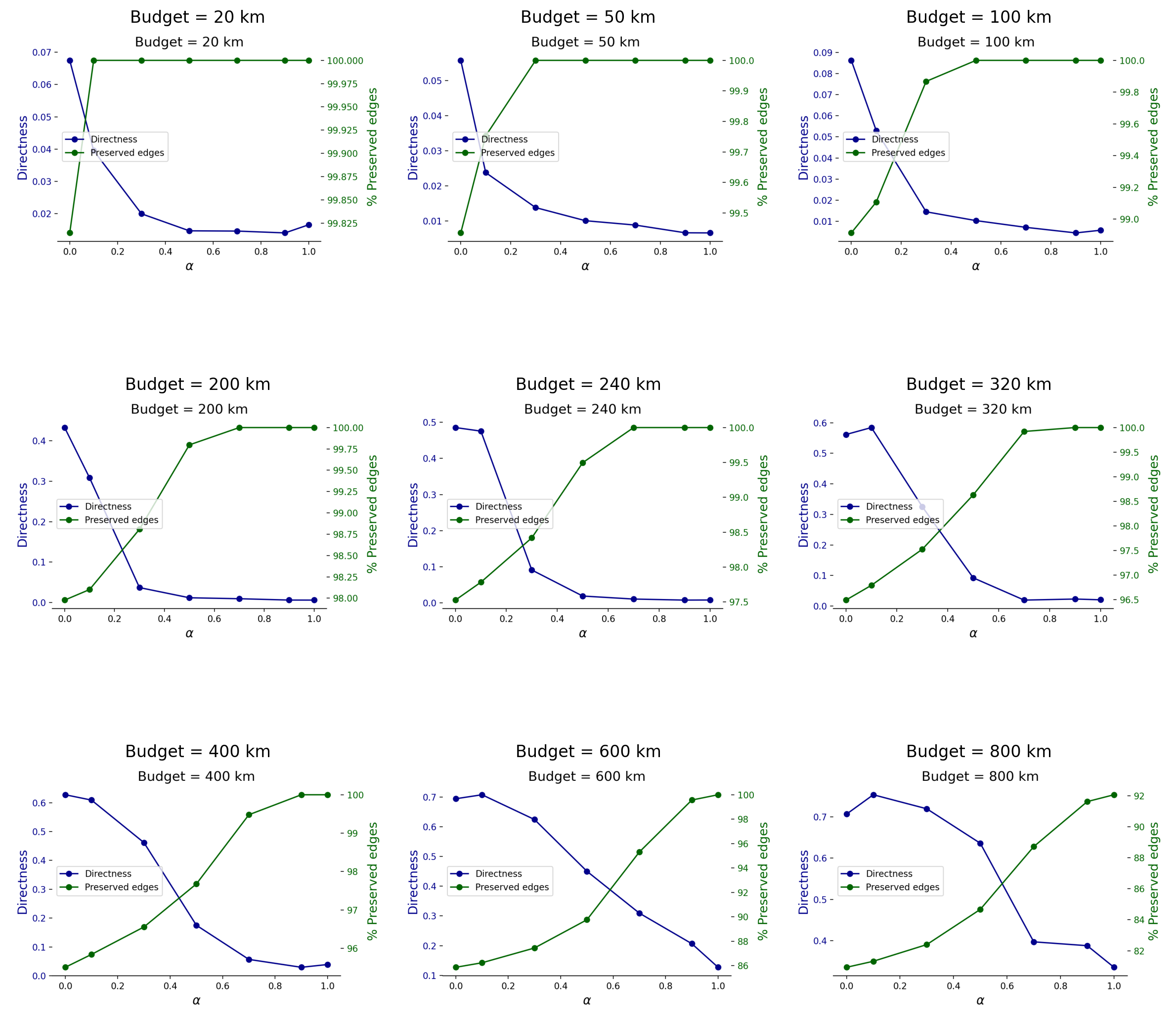}
    \caption{Directness vs Preserved car lanes for many budgets. At lower budgets, lower values of $\alpha$ yield the best sweet spot between preserved lanes and network directness. This is because at low budgets there are enough streets with high betweenness that are still wide enough to not transform too many edges from two-way to one-way.}
    \label{fig:directness vs impact all budgets}
\end{figure}

Figure \ref{fig:directness vs impact all budgets} shows directness and preserved car lanes vs $\alpha$ for selected budgets. At all budgets the general behaviour is the same: there is a tradeoff between the directness of the network and the amount of two way car lanes preserved. Depending on budget availability one can decide which $\alpha$ is suitable for their requirements.

\newpage
	\bibliographystyle{unsrtnat}
	\bibliography{bibliography}